\documentclass {article}
\begin{document}
\begin{center}
{\large\bfseries Observations of Comet 9P/Tempel 1 with the Keck 1
HIRES Instrument During Deep Impact\footnote{The data presented herein were
obtained at the W. M. Keck Observatory, which is operated as a scientific
partnership among the California Institute of Technology, the University of
Califronia and the National Aeronautics and Space Administration.  The
Observatory was made possible by the generous financial support of the W. M.
Keck Foundation.}} \\ [10pt]
Anita L. Cochran \\
McDonald Observatory \\
University of Texas at Austin \\ [5pt]
William M. Jackson \\
Chemistry Department \\
University of California, Davis \\ [5pt]
Karen J. Meech \\
Institute for Astronomy and the NASA Astrobiology Institute \\
University of Hawaii \\ [5pt]
Micah Glaz \\
Chemistry Department \\
University of California, Davis \\ [5pt]
Accepted for {\itshape Icarus} \\ [1in]
\end{center}

\begin{center}Abstract\end{center}
We report high-spectral resolution observations of comet 9P/Tempel 1
before, during and after the impact on 4 July 2005 UT of the Deep Impact
spacecraft with the comet. These observations were obtained with the
HIRES instrument on Keck 1.  We observed brightening of both the dust and gas,
but at different rates.  We report the behavior of OH, NH, CN, C$_{3}$, CH, 
NH$_{2}$ and C$_{2}$ gas.  From our observations, we determined a CN outflow
velocity of at least 0.51 km sec$^{-1}$.  The dust color did not change 
substantially.  To date, we see no new species in our spectra, nor do we see
any 
evidence of prompt emission.  From our observations, the interior material
released by the impact looks the same as the material released from the
surface by ambient cometary activity.  However, further processing of the data
may uncover subtle difference in the material that is released as well as
the time evolution of this material.

\noindent
Proposed Running Head:  Keck Observations of Tempel 1 

\newpage
\section{Introduction}
On 4 July 2005 UT, the Deep Impact spacecraft impacted comet 9P/Tempel 1.
The goal of the impact experiment was to expose deep layers of the
comet nucleus to study the chemical composition
of those layers and to ascertain the structural properties of
the nucleus. 

Comets are the remnant building blocks from the outer solar nebula.
The majority of their lifetimes are spent at large heliocentric
distances so they undergo little chemical evolution.  However, even
far from the Sun, the outer layers are altered via radiation
darkening (Strazzula and Johnson 1991).  Once comets enter the inner
Solar System, sublimation
removes outer layers, while mantling covers the exterior with dust.
\nocite {stjo91}

Comet 9P/Tempel 1 is a Jupiter family comet which has made many
passages into the inner Solar System and has not been extremely
far from the Sun in many orbits.  Thus, the expectation was that
there was an altered outer layer to the comet, but that the interior
was pristine.  The outer layer is thought to be too deep
for a small drill carried by a spacecraft to penetrate to pristine
material. In order to reach fresh, unaltered material,
the Deep Impact mission proposed to create a large crater by placing a
small (370kg) spacecraft in the path of the comet, setting up an impact.
The impact occurred on 4 July 2005 at 05:44:36 UT (Earth-received time
of 05:52:02 UT).  The impact speed was 10.3 km/sec resulting in an impact
kinetic energy of 19 GJ (A'Hearn {\it et al.} 2005).  
\nocite{ahdeepimpact}

The impactor was carried by a larger flyby spacecraft which was able
to monitor the impact with high spatial resolution.  However, while the
flyby spacecraft carried an IR spectrometer, 
it was not able to perform optical spectroscopic observations of
the impact,
nor was it able to monitor the comet for many days after impact.
Thus, a ground-based campaign was established in order to monitor
changes in the comet (Meech {\it et al.} 2005). \nocite{medeepimpact}

We observed the impact and its aftermath from the Keck 1 telescope using the
HIRES spectrograph as part of the ground-based campaign.
High resolution spectra could be used to monitor aspects of the impact which
could not be observed at lower spectral resolution.  These include
monitoring isotope ratios to determine whether the ices in the interior
of the comet are isotopically similar to the ices we normally
see subliming from the surface.  Jehin {\it et al.} (2006) reported the
isotopic ratios for $^{12}$C/$^{13}$C and $^{14}$N/$^{15}$N, both
during the first few hours after impact and at times other than
the impact. They showed that the excavated material is isotopically
similar to the normal outgassing. High resolution spectra
allow for detailed examination of the rotational states of the molecules.
This allowed us to determine whether any prompt emissions were
seen or whether the impact produced any rotationally hot lines.
High resolution also allowed for detection of any new, unexpected
species.  High resolution data are more sensitive to very weak lines
than low resolution spectra because higher spectral resolving power disperses
the underlying continuum to a greater extent, resulting in higher 
line-to-continuum contrast (our spectral resolution of $\sim$6 km\,sec$^{-1}$ 
was not sufficient to resolve typical cometary lines with widths of 
$\sim$1 km\,sec$^{-1}$).
\nocite{jedeepimpact}

\section{Observations and Reductions}
The data were obtained with the Keck 1 telescope and HIRES
spectrograph (Vogt 1992) on 30 May 2005 and 4--6 July 2005 UT 
(see table \ref{log} \nocite{vogthires}
for details of the observations). The Keck HIRES operates with three
$2048 \times 4096$
Lincoln Lab CCD detectors with 15~$\mu$m pixels. Typically, the detectors are
binned in the spatial direction, resulting in 1024 spatial pixels of 0.24
arcsec projected on the sky. The spectrograph has two cross dispersing
gratings to preferentially image the blue or the red part of the spectrum.
We used the UV cross disperser (despite the wavelength
covered, the convention is to call the chips blue, green and red and we will
use these names throughout this paper; blue covers 3047--3951\AA, green
covers 3971--4926\AA, and red covers 4977--5894\AA).
The slit was the ``C1" slit, which images $7 \times 0.86$ arcsec
of the sky (or $4570\times562$\,km at the comet's distance).
This results in a nominal resolving power, 
$\lambda/\Delta\lambda = 47,000$ with the slit projecting to 4.8 pixels on
the detector.
The slit was set at the parallactic angle, which means that it rotated
on the sky during 
the night.  The slit orientation is listed in table~\ref{log}.  By 
setting the slit at the parallactic angle, we ensured that 
atmospheric dispersion changed the position of the comet along the
slit, minimizing loss of light on the slit.  However, it did mean that
we sampled slightly different coma regions throughout the night.  
Thus, we could be sampling overdense regions differently at different
times.

On all nights, bias frames, flat field and ThAr lamps were observed for
calibration (darks were not obtained because these chips typically have
almost no dark current; however, as we will discuss below, this turned out
not to be the case in July).  During the July observations we also observed
a solar analogue
star (HD159222 on 4 July (Hardorp 1978) and 18 Sco (Porto de Mello 1997)
on 5 and 6 July) to be able
to remove the dust continuum and a rapidly rotating hot star (HR 7279
on 4 July and HR 7415 on 5 and 6 July) to be able to remove telluric
features.  On 30 May and 4 July a flux standard was observed.
\nocite{podmesi97,hardorp78}

On 30 May 2005, three 20-minute spectra were obtained as part
of the Keck Director's time.   These observations were intended
to give a pre-impact baseline of how the spectrum of the
coma looked in its normal state.

The observations of 4 July 2005 started during nautical
twilight, at 05:36:15, so that we could obtain a pre-impact spectrum.
Eight degree twilight was at 05:29UT; 12 degree twilight (nautical) was at
05:58UT; 18 degree twilight (astronomical) was at 06:28. 
Since it was not fully dark, the signal
from the sky was substantial in this first spectrum.  The second
spectrum was started at approximately the impact time (start at 05:55:18~UT,
about 3 minutes after impact).  We were
guiding directly on the comet so we could watch the coma brightening.
The exposure time chosen was a compromise between obtaining sufficient
signal/noise and keeping as rapid a cadence as possible.
The observations continued until 09:41:10 UT. The exposure
times were lengthened as the comet dimmed again after the impact and
the airmass of the comet increased.

We observed the comet to brighten rapidly within a few minutes of
impact and to gradually fade as we continued to observe.  This
is in accord with other observers (Meech {\it et al.} 2005;
Schleicher {\it et al.} 2006). \nocite{medeepimpact,scbaba06}
We will discuss the trends of what we observed more in the next
section.

The observations of 4 July 2005 were obtained as Keck Director's
time.  The observations of 30 May and 4 July are available to the
public from the Keck archive.  Because of the data reduction issues
discussed below, we suggest that people interested in obtaining data
from the archive download raw and not processed data.

We also obtained observations on 5 and 6 July 2005 UT to follow up on
the behavior of the comet after the impact.  For some of
the spectra on these two nights, we offset the telescope from the
optocenter to be able to probe the distribution of the gas in
the coma.  The offset positions are noted in the log.
The night of 6 July had some clouds, whereas 30 May and 4 and 5 July
were photometric.  Because of the clouds on 6 July, these observations were
used only for the isotopic abundances but not for monitoring the cometary
activity.

Keck Observatory has a data reduction pipeline, MAKEE, which many
observers use.  However, since it assumes that the target is a 
point source and the ends of the slit are sky, it is not appropriate
for the reduction of cometary data.   Note that the reduced versions
of the data in the archive have been reduced with the MAKEE pipeline
so should be used with caution.

We reduced all four nights' data with the {\it echelle} package in {\it IRAF}.
Each chip was split out to a separate file and the chips were processed
in parallel.  Zero and bias correction and trimming were first performed.
The first 7 pixels and pixels greater than 4000 in the spectral direction
are no good for any chip.  All the images were then flat fielded.

While building flat field images for the July data, we discovered an
enhanced background in the spectral images.  This manifests itself as a signal
that is measurable in the inter-order gaps.  At first, we assumed
that this was scattered light in the spectrograph.  However,
inspection of the comet frames showed that this background scaled
with exposure time, as if there was a high dark current.  This effect
was absent in the May data.  We cannot confirm whether the darks were indeed
elevated for our observations since the dewar had been opened for 
unrelated reasons before we realized the extent of the background problem.

Regardless of the reason for the background, we needed to
remove this from all the spectra or it would have added an unwanted
offset to all spectra.  For the green and red chip
this was not difficult.  It could be removed with the
{\it apscatter} routine for the flat fields. 
For the cometary spectra, we fit a surface to the
inter-order region, working in small sections around each order.

Handling the blue chip background was more complicated because the blue-most
orders overlap one another slightly.  We could define the trace
of the spectrum on the chip with a slit which was less tall than
the 7 arcsec slit we were using for targets. However, with no
inter-order gap between some orders in the actual comet spectra,
there was no background to fit a
surface.  Therefore, we built a ``background" image from a
flat obtained with the shorter slit, scaled that image to
the background of the individual cometary spectra and subtracted
it from the comet spectra.
The fact that the background fit to a flat field spectrum was the same
shape as that underneath the cometary spectra argues again for a
dark current-like source, rather than scattered light since the comet and
the flat field are different colors.

Once the data were flat fielded, the spectra for each order were
extracted.  We used variance weighting to collapse
the spectrum spatially.
We defined the extent of the slit on the chip with the
flat field images.  We fit the dispersion curve to the
ThAr lamp spectra.  The typical rms for the wavelength
solution for each chip was 0.004\AA.  The cometary spectra
were Doppler shifted to the spectrograph rest frame using
the orbital information.

In order to determine the behavior of the dust and coma gas as 
a function of time, we first corrected the spectra for
extinction using mean extinction coefficients for Mauna Kea Observatory
(as given on the Keck Telescope web site; Hodapp (personal communication)
reported that the
extinction was normal for Mauna Kea on 4 July and inspection of our solar
analogue spectra confirm this).
This results in spectra as if seen outside the atmosphere.  Then,
since we obtained the spectra with a variety of exposure times,
each spectrum was normalized to the equivalent count rate for a
15 minute exposure (our most common exposure time on 4 July).  

Cometary spectra consist of three component parts: the emission spectrum
of the gaseous coma, mostly as the result of resonance fluorescence;
the continuous spectrum which underlies the emission that results
from the solar light being reflected from the dust; the telluric
spectrum being superposed on the comet spectrum. The telluric
spectrum only shows up in some orders.  

In order to study the gaseous component, we used the spectra of
the solar analogue stars as a proxy for the solar spectrum and
shifted these spectra to align the absorption features with the
cometary absorption features.  The stellar spectra were scaled
to match the continuum of the comet spectra and then subtracted.
The process was done interactively, with the aim of having
no features left in the continuum regions.
This shift and scale process was performed on each order 
independently.

There were several problems which we encountered with this process.
For some of the orders, especially the bluest orders and observations
at high airmass, the continuum in the comet spectra was very weak, so
it was difficult to align the stellar and cometary spectra.  In those
cases we relied on shifts derived at lower airmass or redder
wavelengths, assuming that all spectra in a given spectral image and night
had the same shift.  This
assumption is accurate to the nearest pixel but not to sub-pixel
values of the shifts.  However, since in these cases the solar
continuum contributed little to the cometary spectra, incorrect
shifts at this level introduce very little additional noise.

For the first spectrum on 4 July, which was obtained with a bright
sky, we actually had two continuum spectra imposed on each gas
spectrum.  The first was the cometary.  The second was the Earth's blue
sky.  The sky spectrum dominated the signal and was not at the
same Doppler shift as the cometary spectrum.  Thus, when removing
the continuum, we preferentially removed the sky spectrum but
would end up with a noisy comet spectrum from the slight wavelength
offset between the comet and sky spectra. 

The most important problem which we encountered for interpretation of
what happened during the aftermath of the impact was that the
continuum brightened much more and faster than did the gas.  For most
orders, this was just taken into account by the scaling process.
However, we had problems removing the continuum for the spectral
order which contains the CN $\Delta v = 0$ band.  In that order,
in the first few spectra where the continuum was strongest, we
found that if we used the criterion that there should be no
absorptions left between the P and R branches, it would appear
that we had over-subtracted continuum to the blue and red of the
band and had imposed ``emission" features.  If we ended up with
a smooth continuum to the blue and (especially) red, then we ended
up with absorptions left between the R and P branches.  We
compromised and ended up with some absorption between the branches.
This effect is illustrated in Fig.~\ref{cnprob}
and most affected spectra 67--70.  

In general, the continuum was strongest in spectra 67--70 and these
spectra had relatively short exposure times.  Thus, these are
the spectra for which the spectral regions between the lines are
noisiest and therefore, these are the spectra with the lowest
signal/noise.

Once the continuum has been removed, the resultant spectra represent
the pure gaseous emission from the comet.  We can subtract these
gas spectra from the original spectra, yielding pure continuum spectra.
These continuum spectra give a measure of the dust.

\section{Results}

\subsection{Dust}
Once we have separate dust and gas spectra we can investigate
the trend of each with time after the impact.  Since there are
no distinct dust features, we maximized the signal/noise by
averaging the dust spectra by chip.  In practice, to avoid
edge effects, we averaged the middle 2000 pixels in wavelength
and all except the first and last order on each chip.
For each spectral image, this results in three average values
centered at $\sim3460$, $\sim4475$, and $\sim5425$\AA\ (the 
``bandpasses" are approximately 750\AA\ wide).  
The upper panel of Fig.~\ref{dustplot} shows the trend of the dust in the
three chips as a function of time.  Note that the average counts
peaked approximately 2300\,sec after the impact and then declined.
The peak is less distinct in the blue chip data than the green
and red chips.  This may or may not be significant since the signal
is so much lower at these blue wavelengths.
The error bars on these data are based on Poisson statistics of the
average counts.  However, since each average is computed from 18,000 --
38,000 pixels, these error bars are extremely conservative.

The initial rise in the average counts indicates that the impact
released a large amount of material, increasing the scattering
cross-section.  This brightening can come from both an increase
in the number of particles, as well as from a change in particle
size distribution to a larger fraction of smaller particles.  
A change in particle size distribution
could potentially result in a change in dust color, depending
on the particle sizes relative to the wavelength of observation.

We can look at the color evolution of the dust by ratioing
the average counts of the different chips. 
These colors are shown in the lower panel of Fig.~\ref{dustplot}.
We see no change in the color of the dust during our observations
of 4 July (note that the 4 July observations span a large range of airmass
so the lack of a color change corroborates the reasonableness of the
adopted extinction coefficients).
We also show the average values for 30 May and 5 July for both the individual
chip counts and for the colors (averaging only those observations
when the slit was centered on the optocenter).  
The colors are similar on all three nights, though the 
4 July data may be slightly redder. 
However, within the error bars, the dust is the same color on 4 July
as on 30 May or 5 July.
This would be in contrast
with the Spitzer observations, where they report a ``bluening"
of the scattered light (Lisse {\it et al.} 2006),
and with Schleicher {\it et al.} (2006), who noted that the material
was redder in color than the general inner coma in the first
15 minutes.  \nocite{lilpsc06}

The average counts for each chip show that the continuum count
rate was lower on 5 July than when we stopped
obtaining spectra on 4 July.  
The 5 July count rate was similar to the 30 May count rate so probably
represents the ambient level.

If the particles are around  0.5--1$\mu$m in size,
as observed by Sugita {\it et al.} (2005), or even in the 0.5--2.5$\mu$m
size range as observed by Schleicher {\it et al.} (2005), we would have seen
a change in color if the particles changed size substantially.
Since we do not see a change in color, there was not substantial
fragmentation or sublimation of the dust in these first two days.
\nocite{sudeepimpact, scbaba06}

\subsection{Gas}
The observations spanned a large wavelength range so we were able
to observe many molecular emission features.  These included
OH, NH, CN, C$_{3}$, CH, C$_{2}$ and NH$_{2}$.  All of these
species are radicals and are believed to be either daughter or
granddaughter species.  The emission spectra are formed
via resonance fluorescence. 
When the impact occurred, some volatile and non-volatile
material was thrown out of the crater.
Keller {\it et al.} (2005) argue that the impact
energy was insufficient, by a factor of several hundred (K\"{u}ppers
{\it et al}. 2005) to sublimate the observed amount of water
molecules, so there was unlikely to be vapor released because
of the impact. \nocite{kuetal2005}
Then, the ices needed to sublime to form
the parent gases and finally they dissociated to form the daughter
and granddaughter species.  Each dissociation step has a
characteristic lifetime, which is typically longer than the time
for the material to flow out of our, rather small ($0.86\times7$ arcsec
or 562$\times$4570\,km),
aperture.  However, some of any parent will be 
dissociated instantly, while other molecules of that parent will take 
more time, so that the lifetime is an e-folding time for the dissociation
process to take place.  The signal which we observe
is that which is imposed upon the ambient spectrum of the cometary coma had
there been no impact.  

We can achieve the highest signal/noise for extracted spectra by summing
the whole length of the 7\,arcsec slit. However, we then must try to deconvolve
the effects of outflow from the effects of the impact.  
In order to understand better the impact effects, we extracted the
spectra only over the spatial resolution of the telescope -- the
inner 0.7\,arcsec (457 km) or 3 pixels in the spatial direction (the
slit width of 0.86 arcsec or 562 km still defines the spectral
resolution of the observations).  These spectra from the inner
pixels still incorporate some of the coma at large distances
along the line of sight but the relative contribution of the near
nucleus region to the coma is increased over spectra extracted
along the full length of the slit. As would be expected, these spectra are
considerably fainter than whole slit spectra (a 10 meter telescope does have its limitations!)
and have lower signal/noise.  In addition, we are prone to the
effects of noise spikes (radiation events) in the extracted spectra
since the cleaning algorithm for such spikes is not effective when
only sampling 3 pixels. We had to edit out noise spikes by hand and
some of the smaller spikes remain (unfortunately not all the noise
spikes are single pixels since radiation events can strike the chip
obliquely). 
Fig.~\ref{spectra-3pix} shows pieces
of the spectra for each of the species listed above.  In this
figure, we show the first spectrum after the impact (file 66),
another spectrum towards the end of 4 July (file 75) and a
spectrum at the start of 5 July (file 129).

From inspection of Fig.~\ref{spectra-3pix} some trends emerge.
First, the comet was not very gaseous to begin with.
If we difference the last pre-impact spectrum (number 65) from
the first post-impact spectrum (number 66), we find there are no
additional emissions seen over the ambient cometary spectrum. Thus,
in the first 10 minutes, no increased emission was detected and the left-hand
panel of Fig.~\ref{spectra-3pix} for each molecule shows, in essence, the
ambient cometary spectrum.

All of the species show an increase in line strength in response
to the impact.  However, the relaxation timescale back to the 
ambient spectrum was different for different species. 
The CN increases dramatically in strength from the impact
to the spectrum 2 hours later, but by 5 July it is back to its pre-impact
level.  Note, however, that the rotational structure of the emissions
does not change due to the impact.  This is true for all species and
argues that there is no prompt emission component due to the impact
and that all species continue to be excited only by resonance fluorescence.

The OH spectra shown here are extremely noisy.  This is especially true
of the spectrum at 08:16 on 4 July because of the large airmass.  While we
corrected for the extinction, these emissions would be the ones {\it most}
affected by any errors in the extinction coefficients applied.  Also, if the
extinction is severe enough that most photons are blocked before they reach
the detector, no correction will repair the signal.  We will discuss OH
more when we discuss full slit extractions below.

C$_{3}$ shows an increase in brightness on 5 July from the
first spectrum on 4 July but this perceived increase may be 
due to the noise in spectrum 66 (recall that this spectrum was obtained
in the first 10 minutes after the impact). 

Inspection of Fig.~\ref{spectra-3pix} shows
that the C$_{2}$ was essentially unchanged (except for the noise level) from the end
of 4 July to the beginning of 5 July (note that the strong line
at around 5155\AA\ is an NH$_{2}$ line, not C$_{2}$).  More can be said
about the C$_2$ if we can increase the signal/noise by summing along
the whole slit.  This is discussed below.

NH is the only molecule
which is essentially as bright on 5 July as it was near the end
of 4 July.  The NH$_2$, however, shows a decay back to the ambient
level by 5 July.  If, as we believe, NH$_3$ is the parent of NH$_2$
and the grand-parent of NH, the earlier decay of the NH$_2$ would
be expected.

CH does not appear to be present in either the first or last spectra.
It is, however, present in the spectrum at 08:16 on 4 July.  This molecule's
spectrum
is helped tremendously by extraction along the complete length of the slit.

In order to boost the signal/noise, we next re-extracted the spectra,
collapsing
the spectra from the full length of the slit.  These spectra are shown
in Fig.~\ref{spectra}.  Spectra are shown at the same three times
as in Fig.~\ref{spectra-3pix}.

In these full-slit extracted spectra, we see the same trend for CN
as we did in the inner three pixels.  Recall that spectrum 66, the 
first shown here, is essentially an ambient cometary spectrum.

The OH appears to continue to increase in line strength for the 5 July
spectrum.  This must be viewed with caution because OH is the
molecule which is most affected by the increasing airmass of the
observations on 4 July.  The observation at 08:08~UT was while
the comet was at $\sim1.8$ airmasses and the extinction is
around 1.4 magnitudes/airmass at this wavelength, even at the high
elevation of Mauna Kea.  Small errors in the extinction coefficient
used would have a large effect.  However, it is apparent that the
OH did not decrease much, if at all, by 5 July.

Inspection of Fig.~\ref{spectra} shows that C$_2$ was essentially unchanged
from the end of 4 July to the beginning of 5 July, but is stronger than
the ambient spectrum. Recall that
C$_{2}$ has no dipole moment and thus has no easy relaxation pathway.
This is also why we normally see such high J-levels for C$_{2}$.
All of the apparent ``noise" to the blue of the bandhead is C$_{2}$ lines
which are so weak in the first spectrum that they are not easily
discerned.  

The behavior of C$_{3}$ is intermediate to that of the CN and C$_{2}$.
It is clearly still stronger on 5 July than the first spectrum, but it
is only half as strong as it was at 08:08~UT.

The NH$_{2}$ is just as strong on 5 July as it
was at the end of 4 July when integrated along the whole slit, 
while the NH has decreased in strength.
With the very blue wavelength of the NH emissions, it would
be affected more by extinction than would the NH$_{2}$, which
would argue that it might be even slightly brighter for the
late spectrum on 4 July. Recall that when we examined the spectra
of these two species in just the inner three pixels spatially,
the NH$_{2}$ actually dimmed sooner than NH, as would be
expected if NH$_{3}$ is the dominant parent of NH.  Thus, the reversed
trend we see here must be a function of the dissociation of the species.

CH behaved like the CN, having almost returned to its pre-impact
value by the next night.  It seems to be only slightly stronger
at the start of 5 July than the first observation on 4 July.

In order to follow the trend of the various species with time in more detail,
we integrated select emission lines for each molecule.  In general,
we fit a continuum under the emission lines and added up the
flux above the continuum in each line individually and summed
a series of lines.  For C$_{2}$, it is impossible to
use individual lines so we summed the bandhead.  For C$_{3}$,
we used the Q- and part of the R-branch.  The integrated intensities
are shown in Fig.~\ref{gasplot}.  It is again obvious that not
all species behaved alike.  As with Fig.~\ref{spectra}, the values
shown here are for the spectra integrated along the full length of
the slit.  All of the species are normalized to the value for that
species in spectrum 66, the first post-impact spectrum.  However, remember
that we saw no emission increase in this spectrum so the value of 1 in this
plot represents the ambient cometary emission.

We measured 9 R-branch lines and 10 P-branch lines of CN [the
``lines" are actually unresolved blends of two closely spaced lines e.g.
R$_1$(11) + R$_2$(10)].  After the impact, the CN lines
grew stronger for the first 4500 seconds and then the intensity
of the lines began to decrease.  At their maximum, the integrated
line intensity was 3 times greater than at the time of impact.
Recall that the slit was $0.86 \times 7$ arcsec, which corresponds
to $562 \times 4570$ km at the comet's distance.  If all of the
CN is produced instantaneously and no more is produced after the
impact, then 4500 sec represents the time
for the material to start flowing out of the aperture.  Since the 
impact was a distinct impulse which caused a release of
material from the impact site, what we measured was the increasing
cross section as the material expanded away from the
nucleus.  Eventually, the area which was fluorescing grew larger
than the footprint of our slit on the cometary coma and we
were no longer able to measure all of the material, though our
aperture does still include the material coming directly towards
us.  That point came at approximately 4500 sec after the impact, causing
the diminution of light we observe. 
If the CN was only produced instantaneously, then the diminution
at 4500 sec indicates that the material was flowing outwards
at 0.51~km~sec$^{-1}$.  This is twice as large as the outflow velocity of
0.2~km~sec$^{-1}$ and of 0.23~km~sec$^{-1}$ found for dust by Meech {\it et al} 
and by Schleicher {\it et al.} (2006), respectively.
Keller {\it et al.} (2005) found a dust velocity of $>0.16$~km~sec$^{-1}$, with
the fastest particles moving at $>0.4$~km~sec$^{-1}$. There are no reported
gas outflow velocities. Keller {\it et al}.
assumed a CN outflow velocity of 0.7~km~sec$^{-1}$ at 1au heliocentric
distance or 0.6~km~sec$^{-1}$ at the comet's distance.
\nocite{kedeepimpact05}

At the time of impact, the first materials released are the parent ices
which must first sublime (unless vaporized by the impact) to the vapor
state before they dissociate to the daughter species.
Initially, the CN represents the
dissociation product of only a very small amount of the parent, implying
that a large quantity of the parent was excavated.
As more CN is dissociated, the intensity of the CN increases. 
But this material is still flowing out
of the aperture.  The combination of the initial component and
the delayed component means that the velocity we detect is a lower
limit to the outflow velocity.

The OH, NH and CH all appear to increase until around 4500~sec and
then decrease.  CH and NH were back to their pre-impact levels
by 5 July.  In Fig.~\ref{gasplot}, the OH appears to reach
its pre-impact levels by the end of our observations of 4 July.
However, this is likely not real.  As the airmass increased, the
extinction at OH increased dramatically.  In the UV, extinction
increases by Rayleigh scattering and follows a $\lambda^{-4}$ law.
While we applied an extinction correction to all the data, at the
larger airmasses essentially no cometary light can penetrate
the atmosphere so the correction fails.  In addition, atmospheric
dispersion begins to cause the image of the comet at the wavelength
of OH to be pushed out of the slit so that we are no longer centered
on the optocenter at OH.  The observations of 5 July show
that the OH emission was still about 1.5 times its pre-impact level.

The behavior of the C$_{3}$, C$_{2}$ and NH$_{2}$ emissions is different
than the other four molecules already discussed.  These molecules
show a monotonic increase in their emissions for the whole time
of our observations on 4 July and were only about 20\% below
their maximum levels of 4 July when we resumed observations on 
5 July.  This indicates that they are being produced faster
than they can flow out of the aperture and are not being dissociated
further in the timescale of observations.  As discussed above,
the C$_{2}$ result is expected since relaxation is a slow process for C$_{2}$.
This is similarly true for C$_{3}$. In addition, the very high line
density of the C$_{3}$ bands makes C$_{3}$ easier to see in noisy spectra
because more lines are included per resolution element.
C$_{3}$ and C$_{2}$ are also grand-daughter species, so should take longer
to form in the first place than daughter species.
However, the monotonic increase is unexpected for NH$_{2}$, generally
thought to be a daughter species.

Unfortunately, integrating
the lines in the spectra which are extracted over only the inner three
pixels produced curves such as those seen in Fig.~\ref{gasplot}
which were too noisy to use to study meaningfully any trends.
Within the noise, it looks like all of the molecules followed
more or less the same trends with time after the initial few minutes.
We have carefully inspected the spectra at all times from these 
extractions of the inner three pixels.  Table~\ref{3pixtrends}
summarizes what we see in those spectra, including the times
of peak brightness for each molecule.  

For CN, we note that there
may have been a change in the relative brightness of a few of the lines
relative to the others.  This may be a signature of the Greenstein
effect.  Further comment on this will have to wait until we can
compute a fluorescence model of the molecule.

\section{Summary}

The spatial extent of the Keck HIRES slit allowed us to monitor the 
material as it flowed outwards superposed on the ambient cometary spectrum.
All species will eventually dissociate as they flow outwards.  However, the 
Keck slit is very short and narrow compared with the scale lengths of the 
observed species.  Thus, the spatial profiles are essentially flat along the
slit.  The off-optocenter spectra of 5 and 6 July may allow us to probe
parent-daughter relationships.  This is beyond the scope of this paper.

Prior to the Deep Impact Mission, our model of comets such as Tempel~1
was that they had a mostly pristine inner nucleus with a mantle
of altered material on the outside.  The depth of this outer mantle
was believed to be a few to tens of meters thick.  By crashing
a spacecraft into the nucleus, the intention was to probe the inner
regions of the cometary nucleus to release more primitive materials.

Our observations of the impact using the HIRES spectrograph on the
Keck 1 telescope were intended to monitor the spectral changes.
The impact certainly caused an increase in brightness from the
release of material from the nucleus.  As the ices sublimed and
dissociated, we would have been sensitive to any changes in the
relative strengths of individual lines which would have resulted
if there had been any prompt emission or energy driven disequilibrium.
We saw none of these (with the possible exception of the CN lines
mentioned above).  All of the molecules brightened and then
dimmed with approximately the same trend.  Thus, we did not even
see a change in relative chemistry as a result of the impact.
Feldman {\it et al.} (2006) observed the impact using the Advanced
Camera for Surveys on HST and reported that CO/H$_{2}$O was 
unchanged by the impact from the ambient, consistent with our 
observations of the molecules in our bandpass.
These results are in contrast with the spacecraft, which saw a change
in the release of organics, HCN and CO$_2$ relative
to H$_{2}$O (A'Hearn {\it et al.} 2005) and Keck 2, which
saw an enrichment in ethane from the impact (Mumma {\it et al.} 2005).
We would not have been sensitive to these species.
\nocite{mudeepimpact2005,fedeepimpact2006}

We have summed all of the spectra in the first few hours after
the impact in order to increase the signal/noise to look for any
new species which were not apparent in the pre-impact spectra
of 30 May.  None are readily apparent.  We plan to investigate
this further by extracting spectra over different cometocentric
distances within our aperture in a future publication.

The impactor succeeded in knocking a large crater in the nucleus,
ejecting $1.5\times10^{32}$ water molecules or $4.4\times10^6$ kg of
H$_{2}$O (Keller {\it et al.} 2005) and
10$^6$~kg of dust (Sugita {\it et al.} 2005).  This should have
exposed fresh material and yet, we observed no chemical changes.
This can only lead to one of two conclusions: 1) The crater was not
deep enough to penetrate the mantle to primitive material, i.e.
the mantle is thicker than we had supposed; or 2) The cometary
material on the outside of the nucleus is not altered significantly
from the interior materials.   Groussin {\it et al.} (2006)
showed that the nucleus has very low thermal inertia. 
Thus, neither the diurnal heat wave or the heat wave from
the extended passage into the inner solar system would 
penetrate deeply into the nucleus.  This would leave pristine
material near the nucleus.  Thus, as we saw with our Keck
data, the interior of the comet did not look substantially
different than the exterior layers and the outer layers
must be very thin.
\nocite{grlpsc06}
\vspace*{0.5in}

\begin{center}Acknowledgements\end{center}
We thank Dr. Fred Chaffee for making Director's Discretionary time available
pre-impact.  Our thanks to Dr. Hien Tran for obtaining the observations
of 30 May and 4 July and for help on 5 and 6 July.
This work was funded by NASA Grant NNG04G162G (ALC), NASA Grant
NNG06A67G (WMJ), NSF Grant CHE-0503765 (WMJ) and by support
from the University of Maryland through a subcontract 2667702 (KJM), which
was awarded under prime contract NASW-00004 from NASA .

\clearpage

\begin{table}
\caption{Log of Observations \label{log}}
\begin{center}
\vspace*{-1.8em}
\begin{tabular}{rccccl}
\hline
\multicolumn{1}{c}{File} & UT & Exposure & Mid & Slit \\
\multicolumn{1}{c}{Number$^1$} & Start & Time (sec) & Airmass & PA$^2$ & Notes$^{3,4}$ \\ \hline \hline
\multicolumn{6}{l}{\bf 30 May 2005 UT} \\
\multicolumn{6}{l}{R = 1.55{\sc au}; $\dot\mathrm R=-4.05$ km/sec; $\Delta$ = 0.75{\sc au}; $\dot\Delta=4.40$ km/sec } \\
1  &    08:33:06.28  &  1200 & 1.19 & 59.3 \\
2  &    08:54:02.14  &  1200 & 1.26 & 63.3 \\
3  &    09:14:55.09  &  1200 & 1.35 & 66.3 \\ [4pt]
\multicolumn{6}{l}{\bf 4 July 2005 UT} \\
\multicolumn{6}{l}{R = 1.51{\sc au}; $\dot\mathrm R=-0.16$ km/sec; $\Delta$ = 0.90{\sc au}; $\dot\Delta=9.22$ km/sec } \\
65 &    05:36:15.26  &  720 & 1.16 & 11.6 & Pre-impact\\
66 &    05:55:17.87  &  600 & 1.18 & 20.1 & Starts right after impact\\
67 &    06:06:12.17  &  600 & 1.19 & 24.8 & Gas first seen above ambient\\
68 &    06:17:05.13  &  900 & 1.21 & 29.7 \\
69 &    06:32:58.93  &  900 & 1.25 & 35.5 \\
70 &    06:48:52.43  &  900 & 1.29 & 40.6 \\
71 &    07:04:46.34  &  900 & 1.34 & 45.0 \\
72 &    07:20:41.49  &  900 & 1.41 & 48.9 \\
73 &    07:36:35.30  &  900 & 1.49 & 52.3 \\
74 &    07:52:29.25  &  900 & 1.59 & 55.2 \\
75 &    08:08:24.56  &  900 & 1.71 & 57.8 \\
76 &    08:24:18.56  &  900 & 1.87 & 60.0 \\
77 &    08:40:13.26  &  1800 & 2.19 & 65.3 \\
78 &    09:11:09.37  &  1800 & 2.87 & 68.6 \\ [4pt]
\multicolumn{6}{l}{\bf 5 July 2005 UT} \\
\multicolumn{6}{l}{R = 1.51{\sc au}; $\dot\mathrm R=-0.04$ km/sec; $\Delta$ = 0.91{\sc au}; $\dot\Delta=9.34$ km/sec } \\
128 &   05:43:50.39  &  900 & 1.17 & 16.2 \\
129 &   05:59:58.55  &  900 & 1.19 & 23.4 \\
130 &   06:16:04.90  &  900 & 1.22 & 29.9 \\
131 &   06:32:02.61  &  1200 & 1.26 & 36.4 \\
132 &   06:53:01.61  &  1200 & 1.33 & 43.0 \\
133 &   07:16:07.57  &  1800 & 1.44 & 50.9 & 17 arcsec South \\
134 &   07:47:36.28  &  1800 & 1.64 & 57.5 & 17 arcsec South \\
135 &   08:18:39.84  &  1800 & 1.93 & 62.5 & 7.8 arcsec South \\
136 &   08:49:38.31  &  1800 & 2.41 & 66.5 & 7.8 arcsec South \\
137 &   09:20:31.67  &  1200 & 3.09 & 67.3 \\ [4pt]
\multicolumn{6}{l}{\bf 6 July 2005 UT} \\
\multicolumn{6}{l}{R = 1.51{\sc au}; $\dot\mathrm R=0.08$ km/sec; $\Delta$ = 0.91{\sc au}; $\dot\Delta=9.46$ km/sec } \\
196 &   06:02:07.03  &  1200 & 1.21 & 25.5 \\
197 &   06:23:01.63  &  1200 & 1.25 & 33.6 \\
198 &   06:43:53.24  &  1200 & 1.31 & 40.5 \\
199 &   07:04:44.35  &  1200 & 1.38 & 46.3 \\
200 &   07:25:39.15  &  1200 & 1.48 & 51.2 \\
201 &   07:46:32.11  &  1200 & 1.61 & 55.3 \\
202 &   08:08:20.87  &  1200 & 1.79 & 58.9 & 17 arcsec North \\
203 &   08:29:16.77  &  1800 & 2.11 & 64.0 & 17 arcsec North \\
204 &   09:00:19.03  &  1800 & 2.73 & 67.6 \\
\hline
\multicolumn{1}{l}{Notes:} \\
\multicolumn{6}{l}{\hspace{1 em}1 Archive file numbers are different than the ones here; use times to convert} \\
\multicolumn{6}{l}{\hspace{1 em}2 Slit Position Angle (PA) at start measured North through East} \\
\multicolumn{6}{l}{\hspace{1 em}3 Slit on optocenter unless noted} \\
\multicolumn{6}{l}{\hspace{1 em}4 Photometric except 6 July (thin clouds)} \\
\end{tabular}
\end{center}
\end{table}

\begin{table}
\caption{Summary of Gas Behavior in Inner 3 Pixels}\label{3pixtrends}
\begin{center}
\begin{tabular}{cp{4.5in}}
\hline
Molecule & \multicolumn{1}{c}{Behavior} \\
\hline
OH & No additional OH until 06:06~UT; Peak intensity at 07:21~UT; returns to original strength by 08:24~UT \\
NH & No additional NH until 06:33 UT; Peak intensity at 07:52; still elevated on 5 July \\
CN & No additional CN until 06:06 UT; Peak intensity at 06:17 UT; may have some change in its rotational structure; returns to original strength by 5 July \\
C$_{3}$ & No additional C$_{3}$ until 06:06 UT; Peak intensity at 06:49 UT and stays at that intensity until 08:24 UT when starts to fade; May be still elevated on 5 July \\
CH & No additional CH until 06:49 UT; Peak intensity at 08:00 UT; Returns to
pre-impact levels by 09:11 UT \\
C$_{2}$ & Possibly no increase in brightness for C$_{2}$; if increases, then it occurs at 08:24 UT \\
NH$_{2}$ & NH$_{2}$ strongly increases in brightness between 05:55 and 06:06 UT; Peak brightness is at 06:33 UT; returns to pre-impact levels by 5 July \\
\hline
\end{tabular}
\end{center}
\end{table}

\clearpage

\section{Figure Captions}

\begin{figure}
\vspace{7in}
\includegraphics{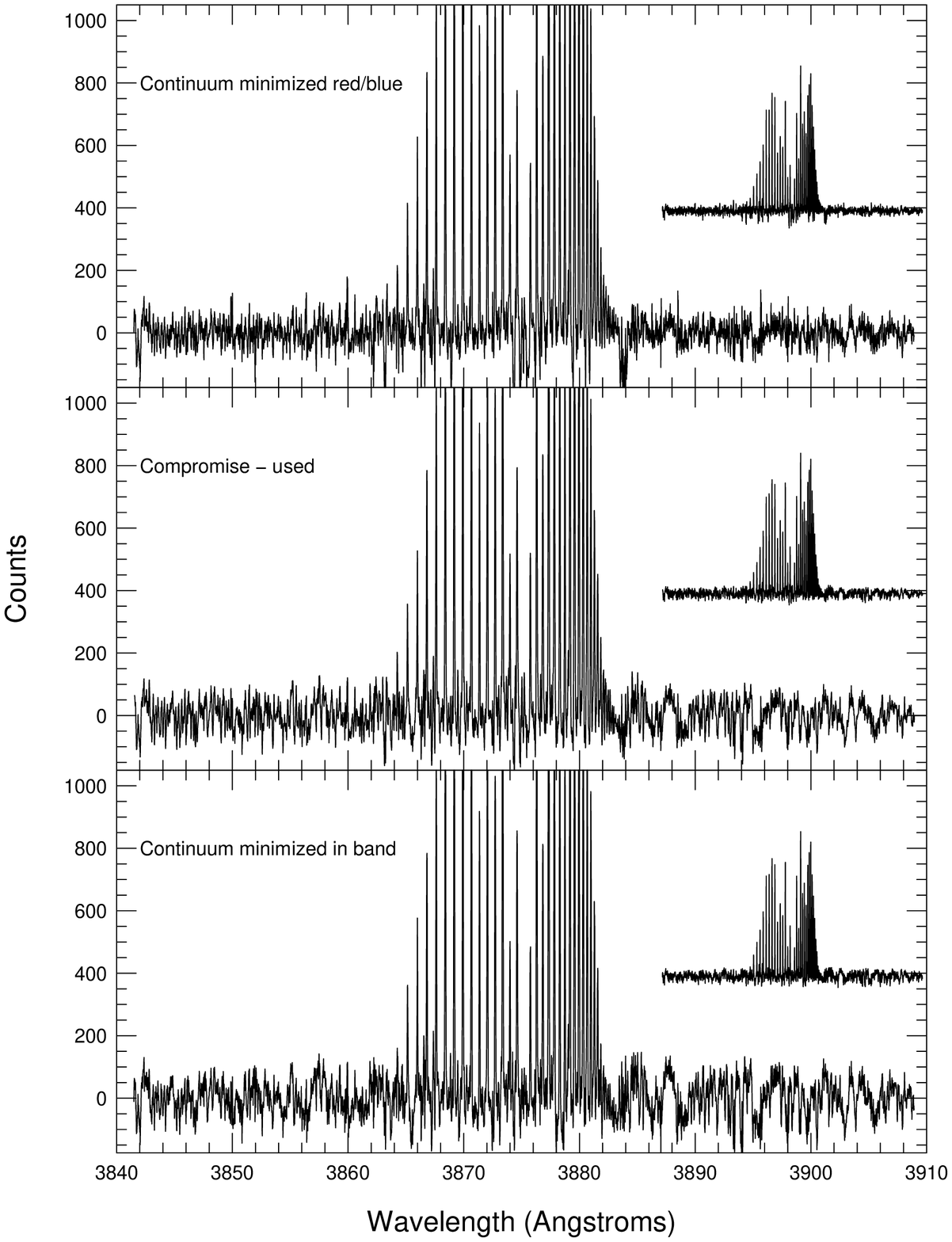}
\caption[fig1]{Cochran {\it et al.}. 2006}\label{cnprob}
\end{figure}

\noindent
{\bf Figure~\ref{cnprob}}:
This figure illustrates the problem with continuum removal under
the CN band from the initial spectra after impact (see text).  This is
illustrated with spectrum 68.  The top panel shows the spectrum when
the solar spectrum is weighted to remove all traces of continuum
to the red and blue of the CN band.  The bottom panel shows
the result when the weighting of the solar spectrum minimizes residual
absorptions within the band.  The middle panel shows the adopted 
compromise spectrum.  In all panels, the complete order spectrum
is shown in miniature as an inset so that the affect the continuum
has on the relative line shapes can be seen.  The line strengths
do not stay exactly the same relative strength.
\vspace{2em}

\begin{figure}
\vspace{7in}
\includegraphics{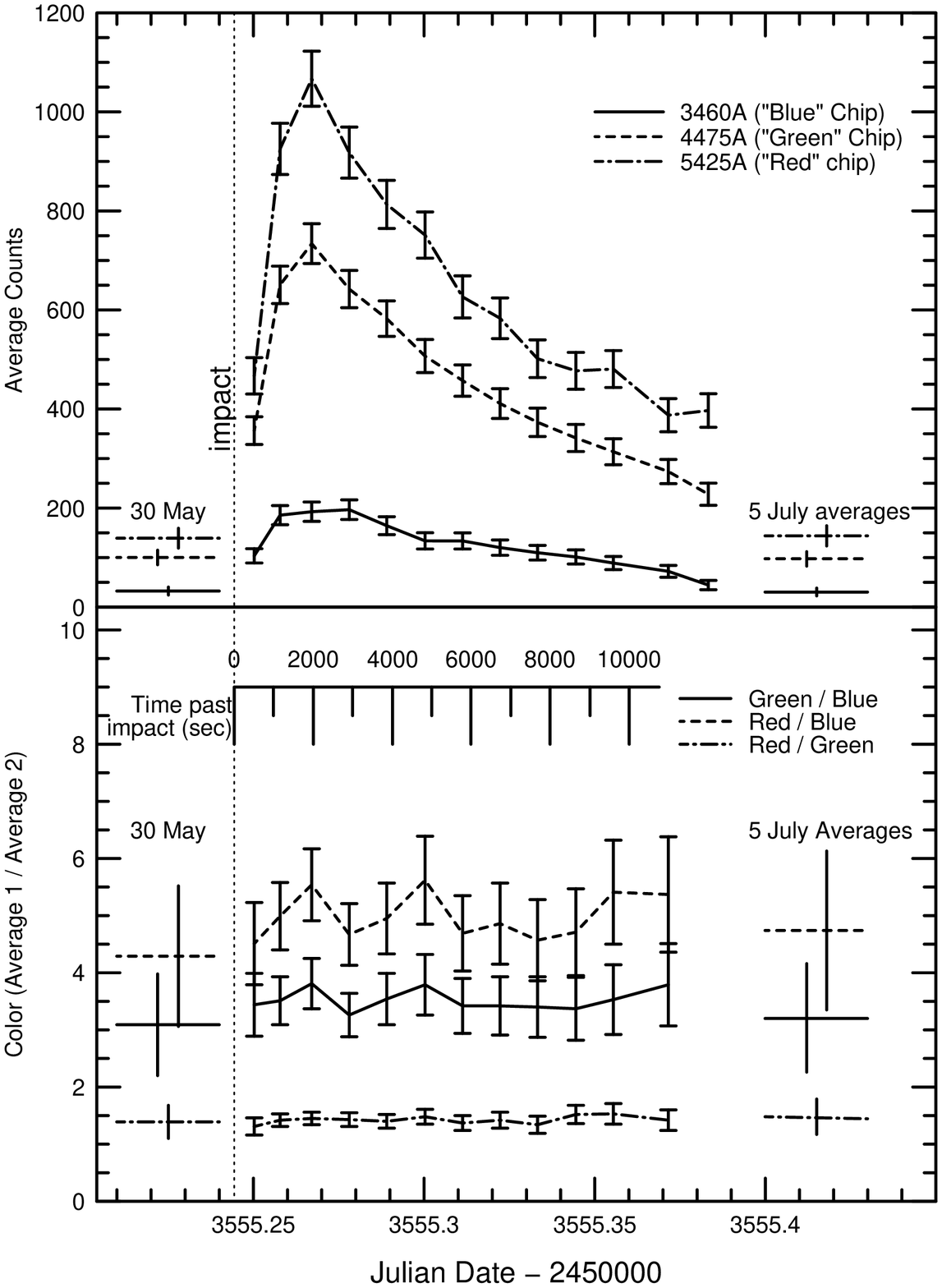}
\caption[fig2]{Cochran {\it et al.}. 2006}\label{dustplot}
\end{figure}

\noindent
{\bf Figure~\ref{dustplot}}: 
The evolution of the dust with time.  Average values were computed
for each spectral image and each chip (see text). The ``blue" chip is centered
at $\sim$3460\AA; the ``green" chip at $\sim$4475\AA; the ``red" chip at
$\sim$5425\AA.  Each bandpass is approximately 750\AA\ wide.  The upper panel
shows the evolution of these average counts with time.  The peak
occurred approximately 2300\, sec after the impact and then the
count rate decayed.  The lower panel shows the color of the dust
with time.  There
is no observed color change.  The average counts from the pre-impact
spectrum on 4 July were not used since the sky was still quite bright,
contaminating this spectral image.  The error bars are extremely
conservative since they do not take into account the large number
of pixels in each average.  For 30 May and 5 July, the error bars
are offset off the center of the bar for clarity.
\vspace{2em}

\begin{figure}
\vspace{7.75in}
\includegraphics{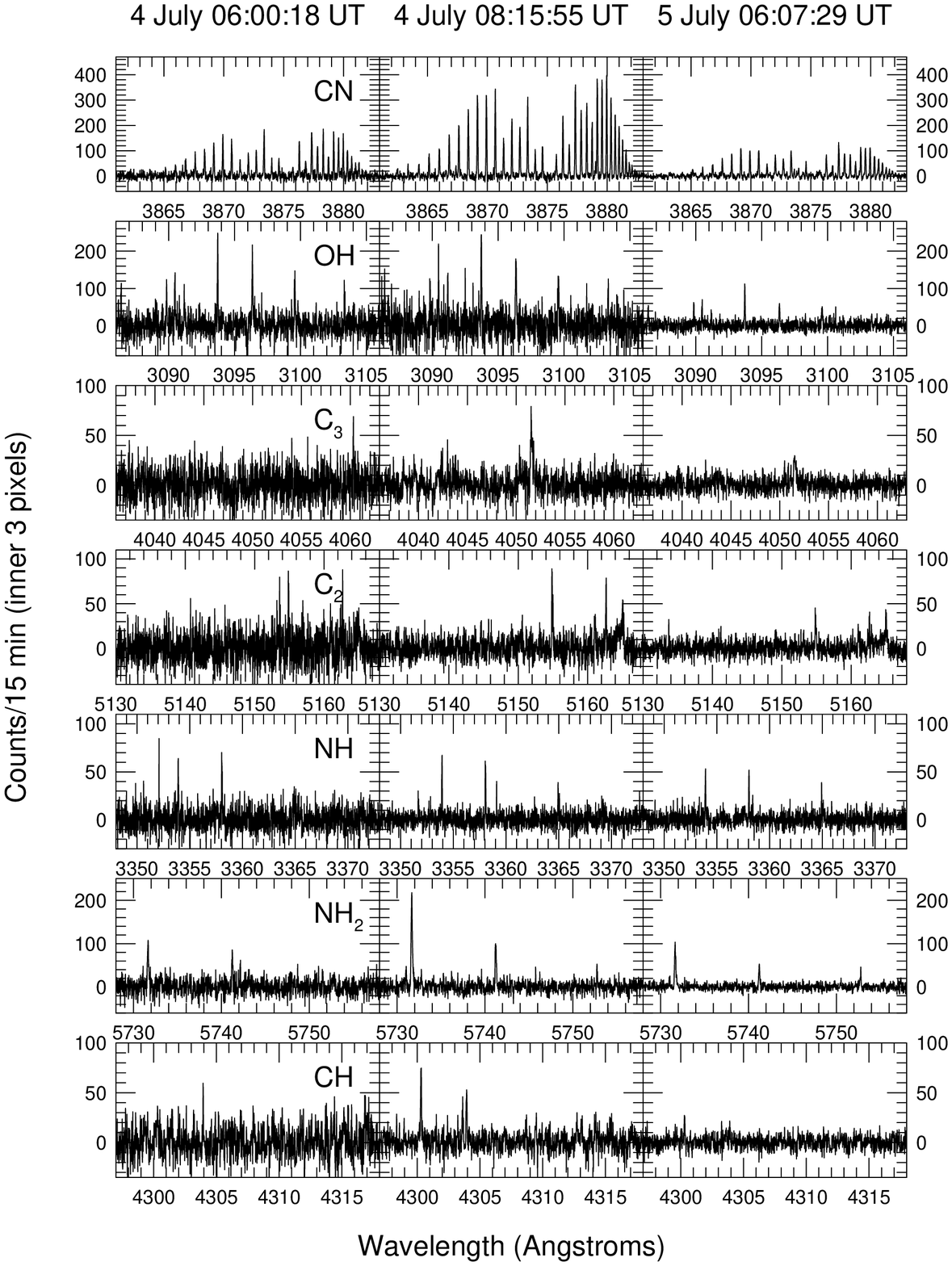}
\caption[fig5]{Cochran {\it et al.}. 2006}\label{spectra-3pix}
\end{figure}

\noindent
{\bf Figure~\ref{spectra-3pix}}:
This figure shows the trend of the molecular emissions when only the
inner three spatial pixels (0.7 arcsec or 457 km) along the slit are considered. 
In this figure we show parts of the spectra at three different times: the
first spectrum after impact, a spectrum towards the end of 4 July
2005, and a spectrum early on 5 July.  The times are noted at the top of each
column and are the mid-times for the observations. 
These three pixels are approximately
the spatial resolution of the telescope so this figure shows the
trend of the gas only at the optocenter and ignores the effects of
gas outflow to first order.  
\vspace{2em}

\begin{figure}
\vspace{7.75in}
\includegraphics{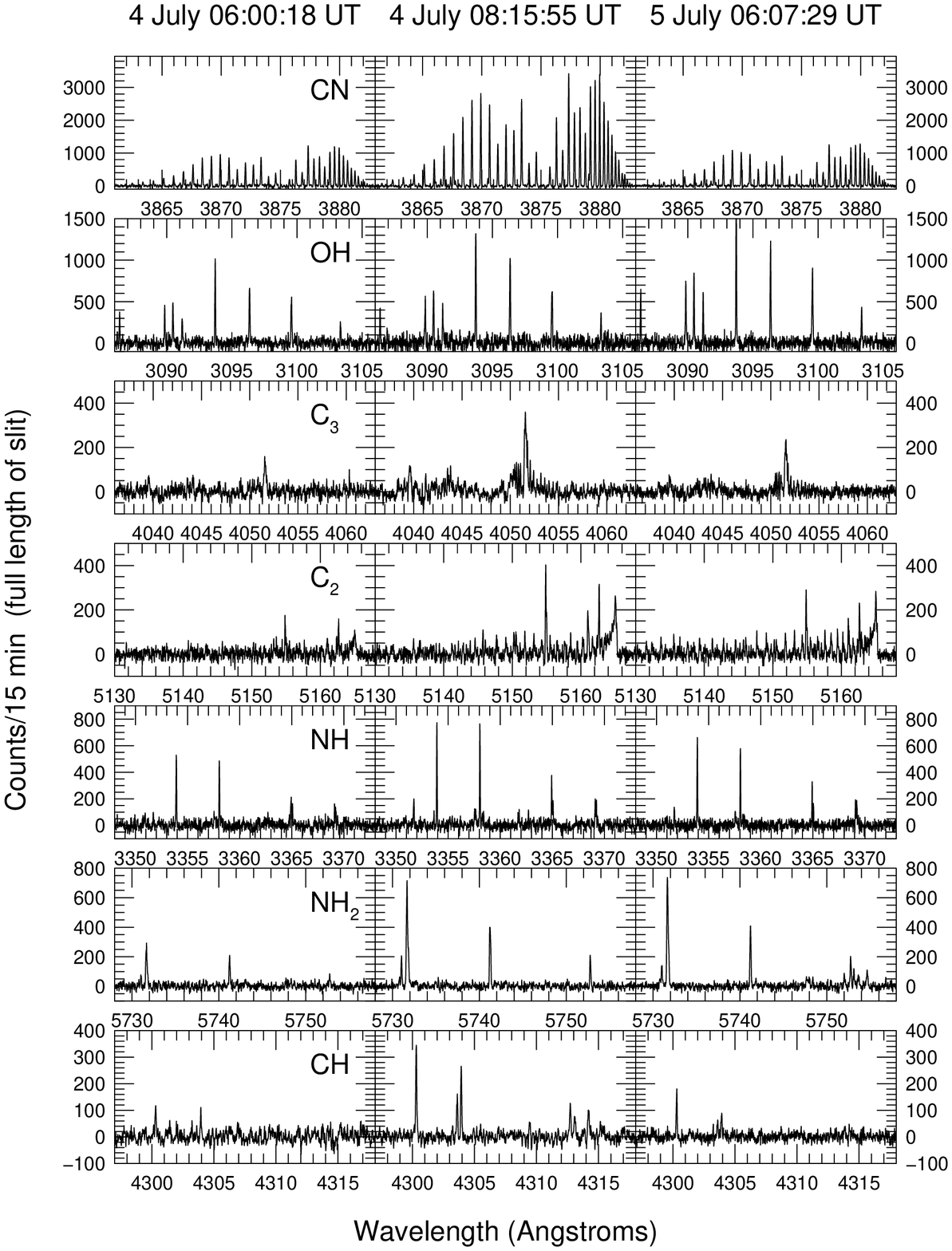}
\caption[fig3]{Cochran {\it et al.}. 2006}\label{spectra}
\end{figure}

\noindent
{\bf Figure~\ref{spectra}}:
This figure shows the trend of the molecular emissions when the spectra are
collapsed along the full length of the 7 arcsec slit.  This figure should
be compared with Fig.~\ref{spectra-3pix}.  The same three
times are shown but the trends are slightly different.
While all line strengths
increased in response to the impact, not all species relaxed on
the same time scale.
\vspace{2em}

\begin{figure}
\vspace{7in}
\includegraphics{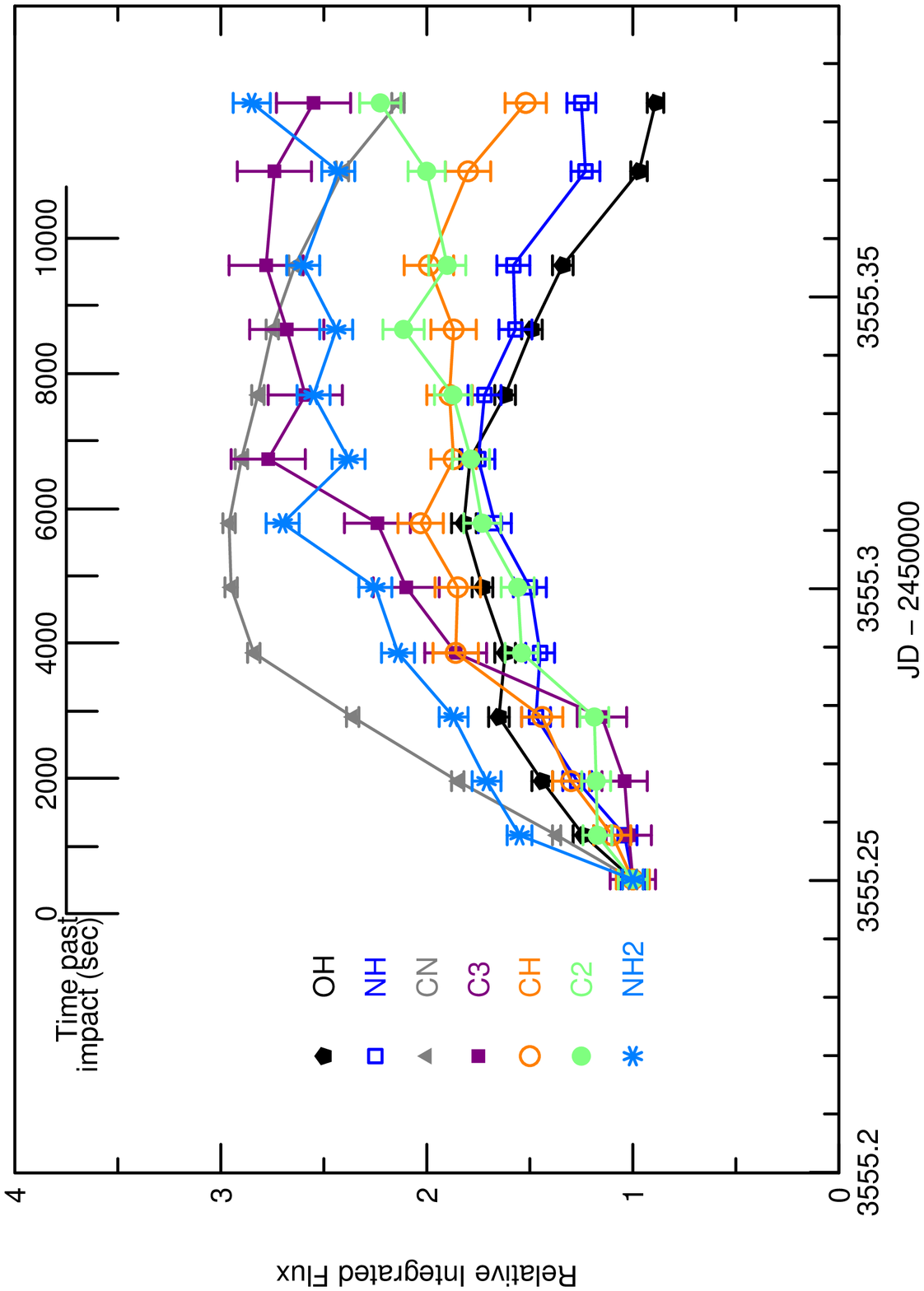}
\caption[fig4]{Cochran {\it et al.}. 2006}\label{gasplot}
\end{figure}

\noindent
{\bf Figure~\ref{gasplot}}:
The evolution of the gas with time.  The counts above the continuum
were integrated along the entire 7\,arcsec slit for various molecular 
emission features and normalized
to the value for spectrum 66, the first spectrum obtained 
after the impact.  Some of these measures include multiple lines
in a dominant band, but for C$_{2}$ and C$_{3}$, we measured
a bandhead. The error bars are the Poisson noise statistics and do not
account for extinction uncertainties, difficulty removing the continuum, etc.
The different molecular species reacted
on different timescales.
Note that we saw no increase in molecular emission in the first
10 minute spectrum after the impact so the value of 1 in this plot
represents the ambient cometary emission and the increase came
after at least 10 minutes post-impact.
\vspace{2em}

\end{document}